\documentclass[useAMS,usenatbib]{mn2e}
\usepackage{lscape}
\usepackage{graphicx}

\title[Using TDG for probing dark matter]{Young tidal dwarf galaxies cannot be used to probe dark matter in galaxies}
\author[Flores et al.]{
H. Flores$^{1},$\thanks{E-mail:hector.flores@obspm.fr}
F. Hammer,$^{1}$
S. Fouquet,$^{2}$
M. Puech,$^{1}$
P. Kroupa,$^{3}$
Y. Yang,$^{1}$
\newauthor
and M. Pawlowski,$^{4}$
\\
$^{1}$ GEPI, Observatoire de Paris, CNRS-UMR8111, Univ. Paris Diderot, 5 place Jules Janssen, F-92195 Meudon, France\\
$^{2}$ Nicolaus Copernicus Astronomical Center, Bartycka 18, PL-00-716 Warsaw, Poland\\
$^{3}$ Helmholtz-Institut fŸr Strahlen- und Kernphysik, Rheinische Friedrich-Wilhelms-UniversitŠt Bonn, Nussallee 14-16, D-53115 Bonn, Germany\\
$^{4}$ Department of Astronomy, Case Western Reserve University, 10900 Euclid Avenue, Cleveland, OH 44106, USA
}
\begin{document}

\date{Accepted; Received; in original form}


\pagerange{\pageref{firstpage}--\pageref{lastpage}} \pubyear{2002}

\maketitle

\label{firstpage}

\begin{abstract}
The location of dark-matter free, tidal dwarf galaxies (TDGs) in the baryonic Tully Fisher (bTF) diagram has been used to test cosmological scenarios, leading to various and controversial results. Using new high-resolution 3D spectroscopic data, we re-investigate the morpho-kinematics of these galaxies to verify whether or not they can be used for such a purpose. We find that the three observed TDGs are kinematically not virialized  and show complex morphologies and kinematics, leading to considerable uncertainties about their intrinsic rotation velocities and their locations on the bTF. Only one TDG can be identify as a (perturbed) rotation disk that it is indeed a sub-component of NGC5291N and that lies at $<$1$\sigma$ from the local bTF relation. 
It results that the presently studied TDGs are young, dynamically forming objects, which are not enough virialized to robustly challenge cosmological scenarios. 
\end{abstract}


\section{Introduction}
The Tully-Fisher (TF, \citealt{Tully77}) relation is probably the most important scaling law of galaxies, relating rotational velocity to luminosity or mass. The baryonic TF (hereafter called bTF) relation is a tight correlation between baryonic mass and rotation velocity \citep{McGaugh00,Verheijen01}  over a considerable range of galaxy velocities, from 20 to 300 km/s \citep{McGaugh10}. There has been considerable debates about the bTF zero point, slope and evolution and whether these can test cosmological scenarios (see, e.g., \citealt{McGaugh05,McGaugh12,Puech10,Dutton11,Kroupa15} and references therein).

Verifying these scenarios can be also done through comparing the bTF locations of dark-matter free galaxies to that of other galaxies. This is because rotation velocities are naturally linked to the total mass encircled by their measurements, and then, the bTF location of a galaxy provides a direct test of its baryonic fraction \citep{McGaugh10}.  Such a test has been firstly attempted by \citet[hereafter B07]{Bournaud07} by deriving the rotation curves of three tidal dwarf galaxy (TDG, dark-matter free), NGC5291N, S and SW, beyond its optical disk. The flat rotation curve was used to deduce the total mass and when compared to the baryonic mass, it reveals the presence of a large fraction of unseen material, which B07 dubbed 'baryonic dark matter'.  \citet{Gentile07} verified the consistency of TDGs with the observed bTF, casting some doubts about the amount of dark matter in the general population of galaxies.

Such a conclusion based on a few objects requires further investigations before being adopted. This has been recently attempted by \cite{Lelli15} who have reinvestigated the measurements of NGC5291N together with that of five other TDGs, providing a significant offset of all TDGs when compared to the bTF of other galaxies. Though the observations of NGC5291N are those of B07, the new modeling causes a doubling of the gas mass, while the B07 method for establishing the inclined-corrected velocity is claimed to overestimate it by a factor 2. Either the latter casts some doubts about the accuracy of such tests, or it requires to better investigate the observed TDGs and verify the overall methodology to capture their locations in the bTF relation.
 
In this paper we reinvestigate the nature of TDGs by using 3D-spectroscopy observations with high spectral resolution and signal to noise. In section 2 we describe the 3D observations and the  high resolution optical images. Section 3 provides full morpho-kinematics analyses  and Section 4 discusses whether or not TDGs can presently be used to test cosmological scenarios. Throughout, we adopt H$_0$ = 70 km/s/Mpc, $\Omega_M$ = 0.3, and $\Omega_\Lambda$ = 0.7.
 
\section{Observations}
\begin{figure*}
\begin{center}
   \includegraphics[width=15cm]{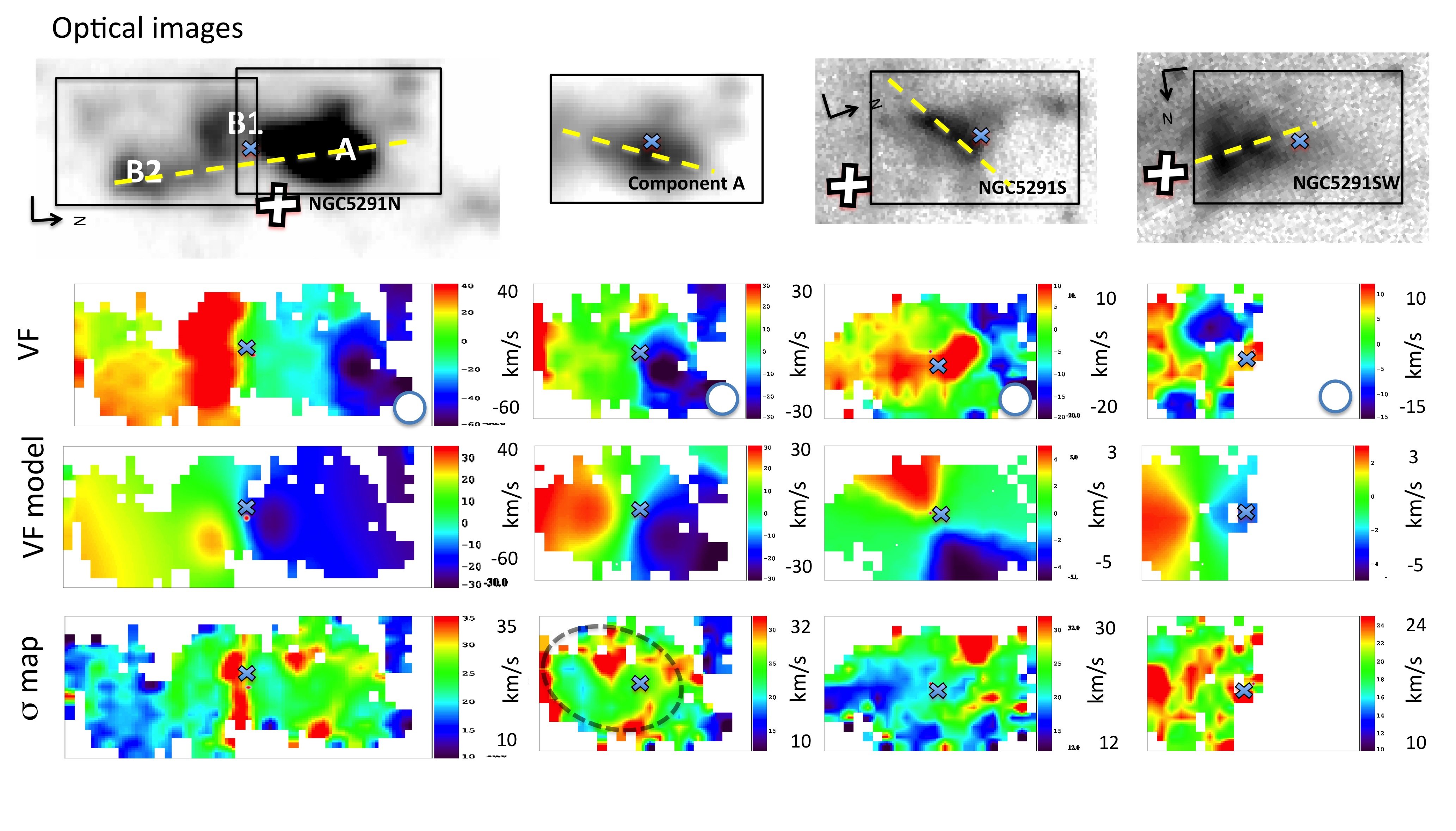}
\caption{{\it Top panels:} Finding charts (R band) of each observed TDG candidate: NGC5291N, the component A, NGC5291S and
NGC5291SW, respectively. In each finding chart, the position of ARGUS FoV is indicated by as a solid rectangle ($\sim$3.2x2.1 kpc$^2$).
The North-East orientation is given (black arrow) as well as the center (small blue cross). The PA is derived from the optical image (dashed yellow line), and the HI center of rotation assumed by \citet{Lelli15} is indicated by a large white cross. {\it Middle-top panels:} Velocity fields from H$\alpha$ that include only spaxels with S/N $\ge$5 and for which we have applied a simple 5 $\times$ 5 linear interpolation for a better visualization, the open circle in the bottom right shows the resolution. {\it Middle-bottom panels:} Rotation models of the  velocity fields assuming the dynamical axis is defined by the stellar disk morphology.  {\it Bottom panels:} Dispersion maps to which we have applied a 5 $\times$ 5 linear interpolation, in the component A a dashed grey line indicates the high [OI]6300/$H\alpha$ distribution from  \citealt{Fensch15}.  }
\label{fig1}
\end{center}
\end{figure*}

We observed three TDGs associated to the system NGC 5291 (N, SW and S) with FLAMES/GIRAFFE (ESO run 093.B-0758(A)). We  used the ARGUS mode (11''x7'' array with 0.52'' width microlenses) with the setup LR6 (R=13700 -- 21 km/s, a resolution 1.5 better than previous Fabry-Perot observations of \citealt{Bournaud04}), and integration times of 3 hours. Observational seeing during 1hr x 3 exposures range from 0.7 to 0.9 arcsec. The data were reduced using the standard ESO pipeline. To test the wavelength calibration, we measured some strong sky lines, showing an error lower than 0.2 km/s. Given the size of TDG NGC 5291 N, we pointed ARGUS in two adjacent regions to construct a larger datacube (final size 41x14 spaxels equivalent to 21".3x7".3, see Fig.~\ref{fig1}), and  combined the data cubes using standard IRAF tasks. 
To construct velocity fields and sigma maps (and S/N maps), we used software developed by \cite{Flores06} and \cite{Yang08}. Each spaxel has been visually checked to avoid sky line and detector residuals (see Flores et al., 2006). Velocity fields and dispersion maps are shown in Fig.~\ref{fig1}. We have also retrieved from ESO archive deep images using VLT/FORS2 (ESO run 82.B-0213(A))
and we have reduced the data using the standard ESO pipeline and the IRAF software. They reveal the morphology of TDG NGC5291N, SW and S, with a seeing of 0.7 and 0.8 arcsec, while the photometry was performed using  Sextractor and Polyphot within IRAF.  Photometric zero points were estimated following the recipe of  FORS2.

\section{Morpho-kinematic analysis of TDGs}
NGC5291 TDGs reveal quite complex and irregular morphologies (see also \citealt{Fensch15}) and are lying within an extended diffuse component, which is part of the NGC5291 tidal debris and are also detected in HI (B07). NGC5291N is even more complex revealing at least three components (we call the brightest ones  A, B1 and B2), for which photometric measurements are presented in Table~\ref{table1}. 
This implies that the single object recorded on the DSS image by B07 is in fact made of at least three objects\footnote{It is indeed a recently formed object ($<$350 Myr; see B07) and could be compared to Fig. 10 of \citet{Kroupa15} for another snapshot of a highly-resolved, young low-mass TDG.}. In Figure~\ref{fig1} we retrieve morphology, velocity field, dispersion map and velocity modeling of the three TDGs (NGC5291 N, S and SW), while the second column presents these properties for the sole, dominant component A of  NGC5291N. 
\begin{table}
\caption{Photometry of the three TDGs of NGC 5291 (N, S and SW), the error associated at each magnitude is $\pm$0.1. Inclination i and PA were determined from R band image and using kinemetry free. Error of stellar  i  and PA is $\pm$ 5 degrees}
{\scriptsize
\begin{tabular}{l c c c c c c}
\hline\hline 
                 & \multicolumn{4}{c}{N} & S& SW \\
                                &  All     & A         & B1     & B2       &             &  \\ \hline
V$_{Bessel}$          & 17.37  & 17.82  &19.64 & 20.21  &  18.63  & 18.18\\
R$_{Special}$         & 17.61  & 17.36  &19.35 & 19.50  &  18.32 & 17.65\\ 
I$_{Bessel}$            & 17.56  & 18.16 &19.50 & 19.93  &  18.03  & 17.52  \\
i$^{1}$                     &  19   &    44& --  &  --     & 30  & 32 \\
PA $^{1}$                 & 102   &  171 & --  &  --     & 113 & -54\\ \hline              
i$^{2}$                      &  16$\pm$10   &    48$\pm$22& --  &  --     & 46$\pm$38  & 32$\pm$29  \\
PA$^{2}$                 & 101$\pm$14   &  163$\pm$8 & --  &  --     &   23$\pm$66   & -46$\pm$58 \\ \hline              
\end{tabular}
}
\label{table1}
Notes: $^{1}$ PA and i from stars; $^{2}$ Assuming PA and i free in the kinemetry.
\end{table}

The  position angle (PA) and ellipticity were determined manually and using the ellipse task (under IRAF). We found a difference of $\sim$20 degrees between the PA of NGC5291N and its component A. Inclinations calculated from R band have been used to recover the rotation curve (RC). The main reason of this choice is dictated by the physics of rotating disks.  In a virialized system, here a rotating disk, stars provide the best, 
indication of the disk center, inclination and PA. \citet[see their Figures 5 to 7]{Lelli15} have provided detailed HI maps of the three systems and have assumed ellipsoidal distribution around the largest velocity gradients, which were assumed to be a rotation. Such a choice leads to rotation centers that are far from the center of the stellar mass distribution (see open crosses in Fig.~\ref{fig1}), which has thus no reason to be interpreted as a rotation disk\footnote{If a disk galaxy forms from gas lying within a uniform sphere
in solid body rotation, the resulting disk has a profile close to
an exponential though slightly more concentrated \citep{Mestel63,Gunn82,Dutton09}. If it is formed through hierarchical
scenario then dissipationless stars are likely aggregating in the center, while viscous, cold gas is more affected by tides \citep{Hopkins08}. Thus stars virialise within a dynamical time and are expected to better indicate the potential of a galaxy in equilibrium, and up to our knowledge, there is no counter- example of a  virialised galaxy having no stars in their center and only having a stellar
population in its outskirts.
}. 

\begin{figure}
\begin{center}
\includegraphics[width=6cm]{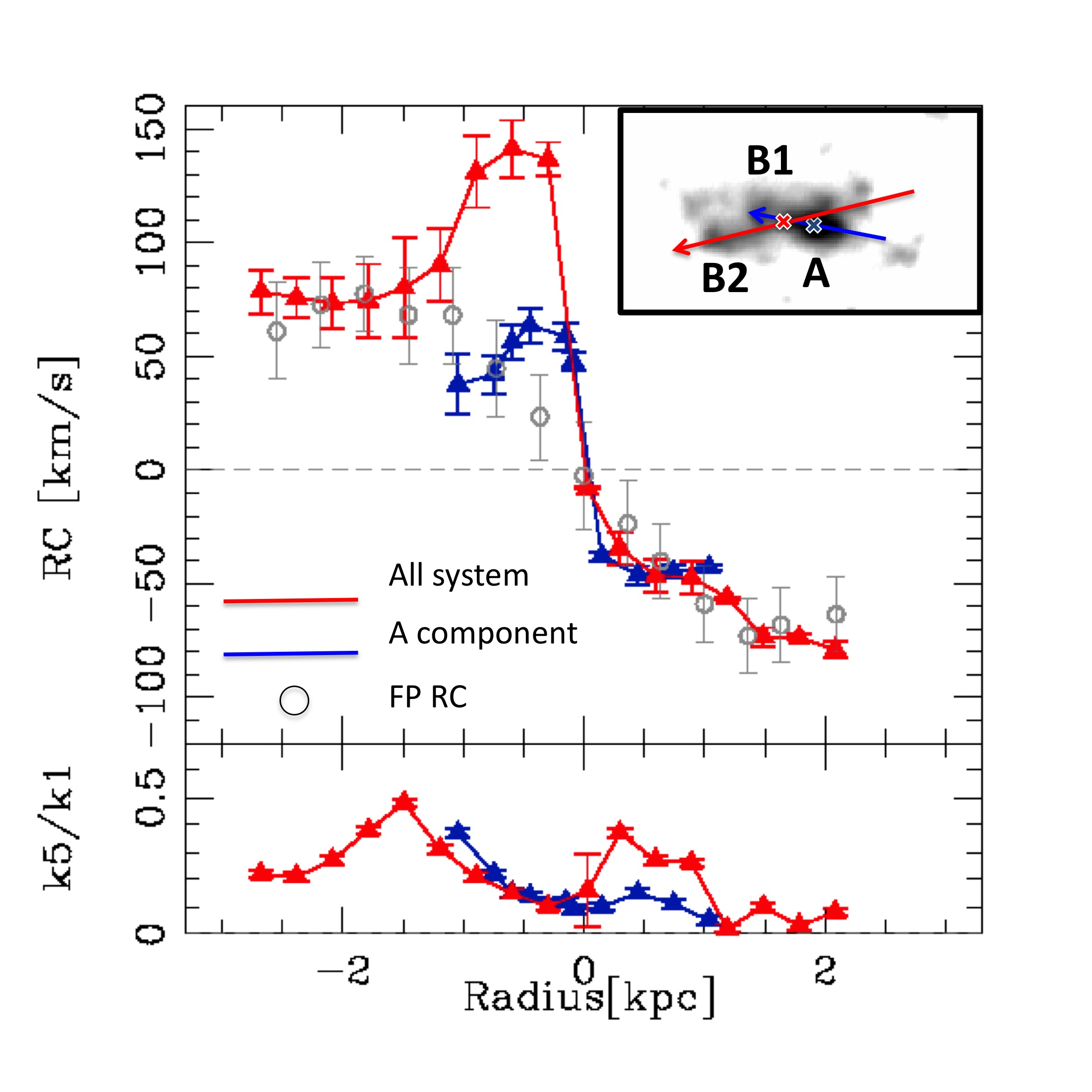}
\caption{{\it Top panel:} Rotation curves with error bars of the whole NGC5291N (red line and triangles) and of  the A component (blue line and triangles). The RC of B07 (open dots) share similarities with that of the whole NGC5291N, although it seems affected by beam smearing effects (see also \citealt{Lelli15}) and it has missed some structures in the velocity field. The insert indicates the adopted PA (red and blue arrows for NGC5291 N and component A, respectively). the last one have the same final $V_{rot}$, the main difference is found in the structures detected in the central regions  of VF. {\it Bottom panel:} k5/k1 ratio along the dynamical axis evidencing that NGC5291N is dominated by non circular motion, while even if the RC of the component A is also asymmetric a $V_{rot}$ can be determined.
}
\label{fig3}
\end{center}
\end{figure}

The 3D H$\alpha$ flux distribution shows peaks at the three main components (A, B1 and B2) plus smaller knots at different velocities. 
The median velocity of B1 and B2 are  63 km/s and 41 km/s  higher  than that of the component A.   This confirms
 that component A can be considered as a single TDG, embedded in a larger gas cloud and will likely merge with the other components in the future, so the final TDG will likely be considerably larger and more massive.  NGC5291N is simply a group of TDGs or of cluster complexes (see, e.g., \citealt{Kroupa15}). Components B1 and B2 show no strong rotation components (full velocity difference $\Delta V$ of 10 and 5 km/s, respectively), which is smaller than the median dispersion (20 and 25 km/s, respectively). Table~\ref{table2} provides the estimates of velocity and $\sigma$ for component A as well as for NGC5291S and SW.
We have analyzed the corresponding maps using the standard software kinemetry \citep{Krajnovic06}, assuming center, inclination and PA from optical images. The kinematics of NGC5291S and SW are too chaotic (see Fig.~\ref{fig1}) to reveal a rotation, even in letting free the dynamical axis. These components are also dominated by quite a large dispersion and by many aspects they are consistent with the complex kinematics class of \citet{Flores06} and \citet{Yang08}. 
Fig.~\ref{fig1} evidences that component A is rotating and Fig.~\ref{fig3} presents its rotation curve together with that of the whole NGC5291N. However component A is not a fully virialized galaxy since it does not show a $\sigma$ peak at the dynamical center. Such a centered $\sigma$ peak is expected for a perfect rotation for which the spatial resolution is not sufficient to fully resolved the velocity gradient between the two maximal velocity plateaus \citep{Flores06,Yang08}. Component A kinematics can be classified as a perturbed rotation and this is corroborated by the importance of non circular motions, for which the relative strength to the rotation is determined by the k5/k1 ratio\footnote{k5 represents the higher-order deviations from simple rotation and points to complex structures on the maps and it is normalized to the rotation term, k1. It is a kinemetric analogue of the photometric term that describes the deviation of the isophote shape from an ellipse (disciness and boxiness, \citealt{Krajnovic06}).} (see Fig.~\ref{fig3}). Values of k5/k1 up to $\sim$ 0.1 can be indicative of a rotating component (see, e.g., \citealt{Krajnovic06}), while values up to $\sim$ 0.5 they are indicative of dominant chaotic motions expected in mergers or close encounters.

There are many clues that the TDGs are indeed forming in a coalescence process, which may not be surprising given the young age of the merger/encounter having occurred in NGC5291 ($\sim$ 350 Myr, B07). The surroundings of component A  show very high values of [OI]6300/$H\alpha$ (see Fig. 7 of \citealt{Fensch15}), which is indicative of shocks. The high [OI]6300/$H\alpha$ distribution  coincides with the $\sigma$ peaks that also form a ring in Fig.~\ref{fig1} (see also Figure 14 of \citealt{Fensch15}). This further suggests that interactions between components are responsible for the shocks, while the shock ring morphology as well as the  absence of a large electron density (see Table~\ref{table2}) are not favoring a strong ionized component linked either to supernova outflow or to hot gas in the NGC5291 halo. The fact that the HI gas peak (see the open cross in Fig.~\ref{fig1} and also figure 5 of \citealt{Lelli15}) is not aligned with stars and ionized gas in component A is also supporting a pre-merger process as in, e.g., \citet[and references therein]{Sengupta15}.  

\begin{table}
\caption{Physical properties deduced from photometry and ARGUS integrated spectra, electron density estimated from $SII[6716]/SII[6731]$, stellar mass from different prescriptions on star ages, the HI mass estimated using HI observations (Lelli et al., 2015), assuming the HI content is associated  to the surface of each stellar component. Masses are given in units of $10^7 M_{\odot}$.}
{\scriptsize
\begin{tabular}{l  c c c| c |c|}
\hline\hline 
                                             & \multicolumn{3}{c}{NGC5291 N} & S & SW \\
                                            &   A   & B1    &  B2   &      & \\ \hline
n$_e$ [cm$^{-3}$]               &  90.0 &26.5   &$<$10.0& 90.7 & 1029.3  \\
M$_* $       &  5.1$^{+5.1}_{-4.6}$  & 1.1$^{+1.1}_{-1.0}$   &  2.0$^{+1.9}_{-1.8}$  & 6.6$^{+6.6}_{-6.3}$ & 19.3$^{+19.3}_{-18.8}$ \\
M$_{HI}$      &  20.0$^{+4.2}_{-4.6}$ &  ---   & ---   & 7.4$^{+2.6}_{-2.5}$  & 6.9$^{+3.2}_{-3.4}$ \\ 
M$_{mol}$    &11$^{+11}_{-7.7}$&  --- &--- & 16$^{+16}_{-8.2}$ &  10$^{+10}_{-3.3}$  \\\hline
M$_{bar}$     &  36$^{+20}_{-17}$  &  ---     &  ---     & 30$^{+25}_{-17}$  &  36$^{+32}_{-25}$  \\
V$_{rot}$$^{1}$          &  40$\pm$10    &  ---     & ---      &  6$^{+3}_{-3}$   &  8$^{+5}_{-4}$ \\
V$_{rot}$$^{2}$         &  28$\pm$10    &  ---     & ---      & 19$^{+14}_{-14}$  &  36$^{+34}_{-34}$  \\
$\sigma$                     &  23$\pm$5     &  ---     &   ---    &  25$\pm$5  & 20$\pm$5 \\ \hline
\end{tabular}
}
\label{table2}
Notes: V$_{rot}$ and $\sigma$ in km/s,  $^{1}$ Assuming PA and i from stars; $^{2}$ Assuming PA and i free in the kinemetry.
\end{table}

\section{The baryonic Tully Fisher diagram}
Studies of galaxy evolution have strikingly shown a considerable evolution of their kinematics even during a relatively recent past (6 Gyr, \citealt{Yang08,Neichel08}). \citet{Flores06} and \citet{Puech08,Puech10} have shown that a significant fraction of these galaxies were strongly offset from the TF and bTF. This is because they are (partly) unvirialized mostly through galaxy interactions and mergers \citep{Hammer05,Hammer09}.  Merger simulations from \cite{Covington10} have shown that many of these systems have locations in the TF biased towards low $V_{rot}$ values. Similar offsets for TDGs are reported by \cite{Lelli15} providing one of the motivations of the present analysis, which reveals that TDG velocity fields are indeed complex (NGC5291S and SW) or show perturbed rotation (component A). 
Table~\ref{table2} presents the mass estimates of the NGC5291 TDGs.  Stellar masses can be calculated from assuming either: (1) that the photometry is dominated by stars coming from the parent galaxies using the \citet{Bell03}  M/L-color relations with I and (V-R) aperture photometry, and, (2) that stellar ages are very young (10 Myr and log(M/L)=-1.6, from \citealt{Bruzual03}) since TDGs are strongly star forming and Wolf Rayet stars have been detected \citep{Duc98}. We consider the stellar mass as being the average between the two above values, keeping the difference as the associated uncertainty. HI masses are evaluated using same apertures than for stars. Large uncertainties are also linked to the molecular gas mass estimates since they are based on a very large PSF (21 arcsec, CO observations from \citealt{Braine01}). We have assumed that molecular masses are ranging from high values (attributing all the CO flux related to the TDG) to low values (performing a scaling based on the aperture ratio between the TDG and the \citealt{Braine01} PSF).

\begin{figure}
\begin{center}
\includegraphics[width=6cm]{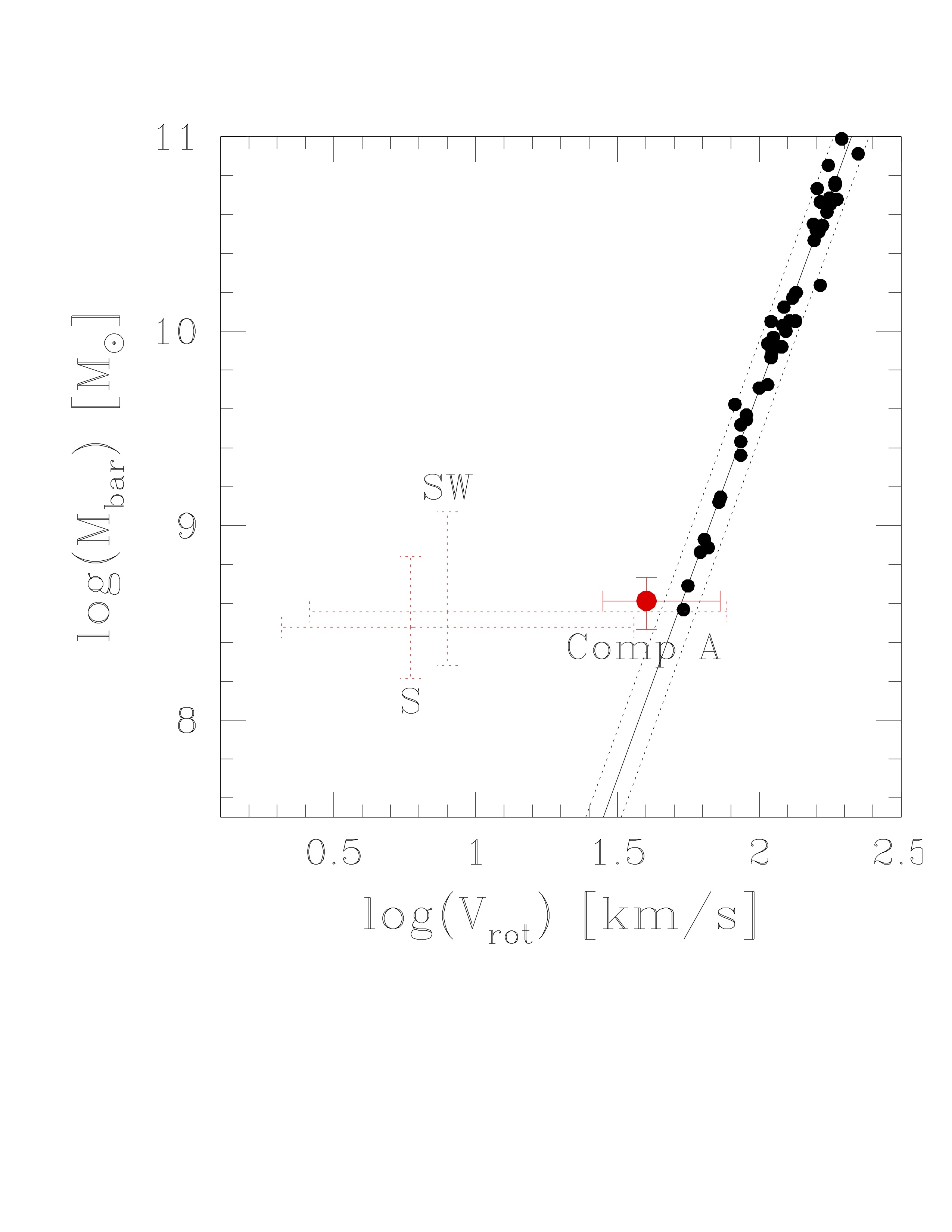}
\caption{Baryonic TF relation of local galaxies from \citet{McGaugh00} and \citet{McGaugh05} together with TDGs and the associated uncertainties. Since NGC5291 S and SW are fully unvirialized we have chosen to represent their points in the right mid-point of their velocity range. Component A appears to be a more mature TDGs and is $<$ 1$\sigma$ from the bTF relation. }
\label{fig4}
\end{center}
\end{figure}

Fig.~\ref{fig4} presents the TDG locations in the baryonic Tully Fisher diagram \citep{McGaugh00,McGaugh05}. Error bars on the NGC5291S and SW velocities are very large since there is no evidence that these complex objects are rotating. Indeed we have assumed their velocities range from the $\Delta$V provided by kinemetry (with letting free the PA and i) to the value provided by a modeling based on the observed optical PA (see Fig.~\ref{fig1} and Table~\ref{table2}).  The semi virialized nature of component A and its well defined dynamical axis lead to much smaller error bars on its rotation velocity. 
It follows that (1) the NGC5291 TDGs are not relaxed nor virialized systems and can not be used to robustly test the dark matter content of galaxies by using their locations in the bTF diagram, and (2) their locations are consistent with the local bTF of galaxies, when accounting for uncertainties linked to their kinematics.

\section{Conclusions}
A thorough analysis of the three TDG candidates in the recent NGC5291 merger illustrates that there is no reason for
them to be interpreted as virialized rotating disk that can be robustly used to test their locations on the bTF relation.
This is similar to many distant galaxies that are suspected to be mergers or merger remnants. \cite{Lelli15}   wondered whether the HI disks are in equilibrium based on the comparison of the disk's orbital time versus the interaction times (B07 modeling). However B07 Figure S7 shows mass distributions that are far more regular than observed.  
 Perhaps this is due to the fact that
B07 modeled the formation of TDGs using a sticky particles scheme for the gas dynamics, 
, which could be problematic for gas-dominated systems as those studied here. 
In the following we will refer to \citet{Ploeckinger15} who elaborated a full hydrodynamic treatment high resolution adaptive particle-mesh method by concentrating on individual TDG, including feedback and incorporating for the first time a treatment of discrete stellar populations via the integrated galactic IMF theory, a more robust representation of TDG formation and evolution. 

 Further attempts to identify their locations in the bTF relation leads to very large error bars for the fully unvirialized systems, while the semi-virialized component A lies close to the bTF of other galaxies (see Fig. 3). We thus conclude that the offsets of the NGC5291 TDGs reported by \cite{Lelli15}  result from a methodological artifact and cannot be straightforwardly used to challenge MOND or another cosmological scenario.

Our conclusion is not changed after examining the three other TDGs studied by  \citet[NGC7252E, NW and VCC2062]{Lelli15}. 
The proposed centres are clearly offset from their stellar distribution, which challenges their interpretation as rotating disks.
It is likely that all these TDGs are observed too shortly after the merger, revealing TDGs in their formation process  and fed by the gas of the tidal tail instead of virialised systems that may result at least one to three billion years later (see, e.g., \citealt{Ploeckinger15} and their Table 2 and Figure 5). 

 However, the B07 test could still have profound consequences on the cosmological scenarios (see, e.g., \citealt{Kroupa10}). This requires to study residuals of much older mergers that could be challenging since tidal tails likely vanish from optical and HI detectability. A possible candidate could be coming from our neighbor, M31, which had probably experienced a major merger, 5.5 Gyr ago \citep{Hammer10}. In modeling the Magellanic System, \cite{Hammer15} have shown that its extent towards IC10 could be a residual of such a tidal tail. Then further kinematic studies of IC10 may be very useful to test its baryonic and dark matter content as well as its location on the bTF. Recall that IC10 lies within the M31 plane of satellites \citep{Ibata13} and is predicted by the M31 modeling to be a TDG lying within the first tidal tail formed in the ancient merger (see Figure 2 of \citealt{Hammer13}). Such studies would also further test the nature of the disk of satellites surrounding the major galaxies, which could be an evidence for a tidal nature of many Local Group dwarfs \citep{Kroupa05,Pawlowski14}.


\begin{thebibliography}{}
\bibitem[Bell et al.(2003)]{Bell03}Bell, E.~F., McIntosh,  D.~H., Katz, N., \& Weinberg, M.~D.\ 2003, ApJ S., 149, 289
\bibitem[Bournaud et al.(2004)]{Bournaud04} Bournaud, F., Duc, P.-A., Amram, P., et al., 2004, A\&A, 425, 813 
\bibitem[Bournaud et al.(2007)]{Bournaud07} Bournaud, F., Duc, P.-A., Brinks, E., et al.\ 2007, Science, 316, 1166 
\bibitem[Bruzual and Charlot (2003)]{Bruzual03} Bruzual, G. and Charlot, S., 2003, MNRAS, 344, 1000
\bibitem[Braine et al.(2001)]{Braine01} Braine, J., Duc, P.-A., Lisenfeld, U., et al.\ 2001, A\&A, 378, 51 
\bibitem[Covington et al.(2010)]{Covington10} Covington, M.~D., Kassin, S.~A., Dutton, A.~A., et al.\ 2010, ApJ, 710, 279 
\bibitem[Duc \& Mirabel(1998)]{Duc98} Duc, P. A. \& Mirabel, F.\ 1998, A\&A, 333, 813
\bibitem[Dutton(2009)]{Dutton09} Dutton, A.~A.\ 2009, MNRAS, 396, 121 
\bibitem[Dutton et al.(2011)]{Dutton11} Dutton, A.~A., van den Bosch, F.~C., Faber, S.~M., et al.\ 2011, MNRAS, 410, 1660 
\bibitem[Fensch et al.(2015)]{Fensch15} Fensch, J., Duc, P.-A.,  Weilbacher, et al.,  2015, arXiv:1509.08873 
\bibitem[Flores et al.(2006)]{Flores06} Flores, H., Hammer, F., Puech et al., 2006, A\&A, 455, 107 
\bibitem[Gentile  et al.(2007)]{Gentile07} Gentile, G., Famaey, B., Combes, F. et al., 2007 A\&A, 472, 25
\bibitem[Gunn(1982)]{Gunn82} Gunn, J. E., 1982, in Br¬uck H. A., Coyne G. V., Longair M. S., eds, AC. PSA, p. 233
\bibitem[Hammer et al.(2005)]{Hammer05} Hammer, F., Flores, H., Elbaz, D., et al.\ 2005, A\&A, 430, 115 
\bibitem[Hammer et al.(2009)]{Hammer09} Hammer, F., Flores, H., Puech, M., et al.\ 2009, A\&A, 507, 1313 
\bibitem[Hammer et al.(2010)]{Hammer10} Hammer, F., Yang, Y.~B., Wang, J.~L., et al.\ 2010, ApJ, 725, 542 
\bibitem[Hammer et al.(2013)]{Hammer13} Hammer, F., Yang, Y., Fouquet, S., et al.\ 2013, MNRAS, 431, 3543 
\bibitem[Hammer et al.(2015)]{Hammer15} Hammer, F., Yang, Y.~B.,  Flores, et al., 2015, arXiv:1510.00096 
\bibitem[Hopkins et al.(2008)]{Hopkins08} Hopkins, P., Cox, T., Hernquist, L.\ 2008, ApJ, 689, 17 
\bibitem[Ibata et al.(2013)]{Ibata13} Ibata, R.~A., Lewis, G.~F., Conn, A.~R., et al.\ 2013,Nature, 493, 62
\bibitem[Krajnovi{\'c} et al.(2006)]{Krajnovic06} Krajnovi{\'c}, D., Cappellari, M., de Zeeuw, P.~T., \& Copin, Y.\ 2006, MNRAS, 366, 787 
\bibitem[Kroupa et al.(2005)]{Kroupa05} Kroupa, P., Theis, C., \& Boily, C.~M.\ 2005, A\&A, 431, 517 
\bibitem[Kroupa et al.(2010)]{Kroupa10} Kroupa, P., Famaey, B., de Boer, K.~S., et al.\ 2010, A\&A, 523, A32 
\bibitem[Kroupa(2015)]{Kroupa15} Kroupa, P.\ 2015, Canadian Journal of Physics, 93, 169 
\bibitem[Lelli et al.(2015)]{Lelli15} Lelli, F., Duc, P.-A., Brinks, E., et al.\ 2015, arXiv:1509.05404 
\bibitem[McGaugh et al.(2000)]{McGaugh00} McGaugh, S.~S., Schombert, J.~M., Bothun, G.~D., \& de Blok, W.~J.~G.\ 2000, ApJ Lett., 533, L99 
 \bibitem[McGaugh(2005)]{McGaugh05} McGaugh, S.~S.\ 2005, Physical Review Letters, 95, 171302 
\bibitem[McGaugh et al.(2010)]{McGaugh10} McGaugh, S.~S.,  Schombert, J.~M., de Blok, W.~J.~G., \& Zagursky, M.~J.\ 2010, ApJ Lett., 708, L14 
\bibitem[McGaugh(2012)]{McGaugh12} McGaugh, S.~S.\ 2012, AJ, 143, 40 
\bibitem[Mestel(1963)]{Mestel63} Mestel L., 1963, MNRAS, 126, 553\bibitem[Neichel et al.(2008)]{Neichel08} Neichel, B., Hammer, F., Puech, M., et al.\ 2008, A\&A, 484, 159 
\bibitem[Pawlowski et al.(2014)]{Pawlowski14} Pawlowski, M.~S.,  Famaey, B., Jerjen, H., et al.\ 2014, MNRAS, 442, 2362 
\bibitem[Ploeckinger et al.(2015)]{Ploeckinger15} Ploeckinger, S., Recchi, S., Hensler, G., \& Kroupa, P.\ 2015, MNRAS, 447, 2512 
\bibitem[Puech et al.(2008)]{Puech08} Puech, M., Flores, H., Hammer, F., et al.\ 2008, A\&A,, 484, 173 
\bibitem[Puech et  al.(2010)]{Puech10} Puech, M., Hammer, F., Flores, H., et al.\ 2010, A\&A, 510, A68 
\bibitem[Sengupta et al.(2015)]{Sengupta15} Sengupta, C., Scott, T.~C., Paudel, S., et al.\ 2015, arXiv:1509.02614 
\bibitem[Tully \& Fisher(1977)]{Tully77} Tully, R. B., \& Fisher, J. R. 1977, A\&A, 54, 661
\bibitem[Verheijen(2001)]{Verheijen01} Verheijen, M.~A.~W.\ 2001, ApJ, 563, 694 
\bibitem[Yang et al.(2008)]{Yang08} Yang, Y., Flores, H., Hammer, F., et al.\ 2008, A\&A, 477, 789 
\bibitem[Yang et al.(2014)]{Yang14} Yang, Y., Hammer, F.,  Fouquet, S., et al.\ 2014, MNRAS, 442, 2419 
 \end{thebibliography}
\end{document}